\documentclass{DISproc}

\usepackage{mcite}

\begin{document}
\title{Heavy Flavour Working Group Summary}

\author{{\slshape Jolanta Brodzicka$^1$, Massimo Corradi$^2$, Ingo
    Schienbein$^3$, Reinhard Schwienhorst$^4$}\\[1ex]
$^1$ Krakow University,
$^2$ INFN Bologna,
$^3$ LPSC Grenoble,
$^4$ Michigan State University}

\contribID{xy}

\doi  

\maketitle

\begin{abstract}
We review theoretical and phenomenological aspects
of heavy flavour production as discussed in the heavy flavour
working group of the DIS 2012.
Recent theoretical progress includes approximate NNLO calculations
for heavy quark structure functions in deep inelastic scattering,
the extension of the ACOT heavy flavour scheme to jet production,
and advances in top physics where the highlight is clearly the
first complete NNLO QCD prediction for top pair production in the
$q \bar{q}$ annihilation channel.
Furthermore, state of the art phenomenological predictions for 
open charm and bottom, charmonium, and single top
and top pair production are discussed in addition to other topics
such as the effect of double parton scattering on heavy quark production.
New measurements on charm and beauty production presented in the heavy
flavor working group are summarized and discussed in comparison with
QCD predictions. Top quark strong and weak couplings 
as well as top quark properties are being measured with precision at the LHC 
and the Tevatron. 
We summarize also recent results on spectroscopy of charmonia, bottomonia
and $b$-hadrons, along with studies of their decays and properties.
Searches for physics beyond Standard Model through precise measurements
of rare decays of heavy flavours are discussed as well.
\end{abstract}

\section{Introduction}

The measurement of heavy quark production provides a test of many
aspect of QCD.  A considerable progress in the QCD calculations for
heavy-quark production has been done in the recent years. The
theoretical results presented at this workshop are reviewed in Section~\ref{sec:th}.
Many new 
results have been presented on
the production of heavy quarks from different types of collisions:
deep inelastic scattering, photoproduction, $pp$ and
$p\bar{p}$, and nuclear collisions (PbPb, AuAu, CuCu, dAu).
Results on charm and beauty production are summarized in
Section~\ref{sec:cb}. Due to space constraints the results from heavy-ion
collisions, which would deserve a dedicated summary, are not
discussed.  The measurements of top quark production are discussed in
Section~\ref{sec:top} with the new results on top properties. Finally
Section~\ref{sec:other} summarizes the updates in heavy hadron
spectroscopy and on searches beyond the Standard Model exploiting
B-hadron decays.
 
\section{Theory}
\label{sec:th}

\subsection{Deep inelastic scattering}
\label{sec:th_dis}
Deep inelastic scattering (DIS) data form the backbone of global analyses
of parton distribution functions (PDFs). Precise determinations of PDFs
require the inclusion of heavy (charm, bottom) quark mass terms in the 
calculation of DIS structure functions at higher orders in perturbation
theory. 
In fixed order perturbation theory the charm contribution $F_a^c$ to the inclusive
DIS structure function $F_a$ ($a=2,L$) is given as a convolution of PDFs
$f_j$ with heavy quark Wilson coefficients $H_j$ ($j=g,u,d,s$):
\begin{equation}
F_a^c(x,Q^2) = H_{a,j}(x,\frac{Q^2}{\mu^2},\frac{m^2}{\mu^2}) \otimes f_j(x,\mu^2)\, ,
\end{equation}
where $m$ is the charm quark mass and $\mu$ the factorization scale (which has
been identified with the renormalization scale).
For neutral current DIS, the coefficients $H_{a,j}$ are currently known to order 
${\cal O}(\alpha_s^2)$.
At this conference, progress has been reported 
to construct approximate heavy quark Wilson coefficients at order 
${\cal O}(\alpha_s^3)$ using the information from different kinematic limits.

Due to mass factorization, in the limit $Q^2 \gg m^2$ the $H_{a,j}$ can be written
as a convolution of light flavor Wilson coefficients $C_{a,i}$ with universal 
operator matrix elements (OMEs) $A_{ij}$:
\begin{equation}
H_{a,j} \simeq C_{a,i}(x,\frac{Q^2}{\mu^2}) \otimes A_{ij}(x,\frac{m^2}{\mu^2})\, .
\end{equation}
Indicating the loop order as upper index one can write generically
\begin{equation}
H^{(3)} \simeq C^{(0)} \otimes A^{(3)} + C^{(1)} \otimes A^{(2)} + C^{(2)} \otimes A^{(1)}\, ,
\end{equation}
where the light flavor Wilson coefficients are available up to order ${\cal O}(\alpha_s^3)$ 
\cite{Vermaseren:2005qc}.
In order to construct the heavy quark Wilson coefficient functions $H^{(3)}_{a,j}$
in the asymptotic limit, the OMEs are needed up to the same order ${\cal O}(\alpha_s^3)$
and partial results for certain color factors have been obtained very recently
\cite{Blumlein,*Blumlein:2012vq,*Ablinger:2012qm}.
As an important application, the OMEs are needed for the definition of variable flavor
number schemes since they enter the matching conditions for the PDFs with $n_f$
and $n_f+1$ active flavours.

In addition to the asymptotic limit $Q^2 \gg m^2$, 
it is possible to exploit universal features in the threshold region
$\beta = \sqrt{1 - 4 m^2/s} \simeq 0$ where Sudakov logarithms $\ln \beta$ 
can be resummed 
and in the high energy limit where information on leading and next-to-leading
small-$x$ logarithms is available in order to construct improved heavy quark 
coefficient functions at order ${\cal O}(\alpha_s^3)$ \cite{Moch,Kawamura:2012cr}.

As is well-known, the heavy quark coefficient functions $H_i$ are not IR-safe:
$H_i \to \infty$ for $\frac{Q^2}{m^2} \to \infty$. Therefore, most of the modern
global analyses of PDFs use so called general mass variable flavor number 
schemes (GM-VFNS) where the large logarithms $\ln \frac{Q^2}{m^2}$ are removed
from the coefficients $H_i$ and resummed by heavy quark initiated subprocesses
with evolved heavy quark parton distributions.
At the same time, finite mass terms $\frac{m^2}{Q^2}$ are retained in the subtracted IR-safe
Wilson coefficients $\hat H_i$.
For use in precision determinations of PDFs, these GM-VFNS have to be formulated
at NNLO and F.\ Olness presented approximate results for the 
neutral current structure functions $F_2$ and $F_L$
in the ACOT GM-VFNS \cite{Aivazis:1993pi} up to order ${\cal O}(\alpha_s^3)$ 
\cite{Stavreva:2012rm,*Stavreva:2012bs}.

\subsection{Open charm and beauty}
\label{sec:th_opencb}
Most of the heavy flavour schemes have been formulated 
for inclusive DIS processes and are used in global analyses of PDFs.
On the other hand, less inclusive processes provide many
additional tests of pQCD in various kinematic regions 
and are closer to the experimental measurements.
However, theoretical calculations are much more challenging
and even more so if heavy quark mass effects have to be taken
into account.
Nevertheless, any heavy flavour scheme should
also be applicable to less inclusive observables in order to be 
considered a general formalism of perturbative QCD (pQCD) including 
heavy quark masses \cite{Thorne:2008xf,*Olness:2008px}. 
Relying on a factorization theorem with heavy quarks \cite{Collins:1998rz},
ACOT-like variants of the GM-VFNS have been applied to inclusive 
heavy meson production in DIS 
\cite{KS},
photoproduction \cite{photo,*Kniehl:2009mh}, 
hadroproduction 
\cite{Olness:1997yc,hadro},
and electron-positron annihilation \cite{Kneesch:2007ey}.
At this conference updated numerical results for inclusive $D$ and $B$ meson production 
in the GM-VFNS at the LHC have been presented 
\cite{Spiesberger:2012zn,*gmvfns_lhc}.
%
Furthermore, a new method for calculating DIS jet production in the ACOT scheme
has been reported extending the dipole subtraction formalism to all possible
QCD splitting processes with heavy quarks including splittings of coloured, massive
particles in the initial state 
\cite{Kotko}.
%

An interesting feature of exclusive processes with a heavy quark
in the final state is that they are useful to probe heavy flavour PDFs.
In most global analyses of PDFs the charm and bottom distributions
are generated ``radiatively'' using perturbatively calculated boundary 
conditions. With other words, no new fit parameters are associated to the
charm and bottom PDFs.
However, a purely perturbative treatment might not be adequate,
in particular for the charm quark with mass $m_c \simeq 1.5$ GeV,
and in fact non-perturbative models exist that predict an
intrinsic charm (IC) component in the nucleon \cite{Pumplin:2007wg}.
Clearly, the heavy quark PDFs should be tested since they play an
important role in some key processes at the LHC \cite{Maltoni:2012pa}.
A promising way to constrain models on IC is
the measurement of inclusive $D$ meson production at RHIC
or with the LHCb detector at the LHC 
where at forward rapidities the differential cross section
can be enhanced by a factor of up to 5 compared to the prediction
with a radiatively generated charm PDF 
\cite{Spiesberger:2012zn,*Kniehl:2012ti,*Kniehl:2011bk,*Kniehl:2009ar}.
Another process which is very sensitive to the heavy quark PDF is direct 
photon production in association with a heavy quark jet \cite{Kovarik:2012pj}.
Data from the D0 experiment at the Tevatron \cite{Abazov:2009de,Abazov:2012ea}
overshoot the standard NLO QCD predictions \cite{Stavreva:2009vi}
at large transverse photon momenta and the inclusion of an intrinsic heavy quark 
component in the nucleon can reduce the difference between data and theory 
but not fully resolve it.
Measurements of this process at RHIC and the LHC probe the heavy quark PDFs
in different regions of the momentum fraction $x$ and would be useful to shed more
light on the current situation. 
In addition, the measurements at the LHC would provide a baseline for $\gamma+Q$ production
in $pA$ \cite{Stavreva:2010mw} and $AA$ collisions \cite{Stavreva:2011wc}.

Finally, work has been presented on $c\bar{c}$ production in the
$k_T$-factorization formalism. The predictions for $D$ meson production
at the LHC somehow undershoot the data of ALICE and LHCb (preliminary).
Furthermore, the production of two $c\bar{c}$ pairs
in a formalism with double-parton scattering has been discussed. 
The predicted cross sections for $c\bar{c}c\bar{c}$ at the LHC 
receive similar contributions from single-parton and from double-parton scattering 
\cite{Szczurek:2012is,*Luszczak:2011zp,*Schafer:2012tf}.

\subsection{Charmonium production}
\label{sec:th_charmonium}
The charmonium $J/\psi$ has been extensively studied experimentally 
ever since its discovery in 1973. However, theoretically, heavy quarkonium
production and decay are still not well understood.
A rigorous framework for theoretical studies is provided by the factorization
theorem of nonrelativistic QCD (NRQCD) \cite{Bodwin:1994jh}
where the charmonium production cross section factorizes into calculable 
short distance cross sections for the production of a heavy quark pair $c\bar{c}[n]$
in a Fock state $n$ and nonperturbative long distance matrix elements (LDMEs)
$\langle {\cal O}^{J/\psi}[n]\rangle$ which have to be extracted from experiment.
The Fock states are
described by quantum numbers for spin, orbital and total angular
momentum and color, $n={^{2s+1}\!L_J^{[c]}}$.
For $J/\psi$ production the following states are considered:
$n=
{^3\!S_1^{[1]}}, 
{^1\!S_0^{[8]}},
{^3\!S_1^{[8]}},
{^3\!P_J^{[1]}}$ 
including color-octet (CO) states $[c]=[8]$
in addition to color singlet (CS) states $[c]=1$.
%

A NLO NRQCD analysis of the $J/\psi$ yield and polarization
based on the results in 
\cite{Butenschoen}
can be summarized as follows \cite{Kniehl,Butenschoen:2012qh}:
(i) A global analysis of unpolarized world $J/\psi$ data from hadroproduction,
photoproduction, two-photon scattering and electron-positron annihilation experiments 
allows to determine the three CO LDMEs  
$\langle{\cal O^{J/\psi}}({^3\!S_1^{[1]}})\rangle, 
\langle{\cal O^{J/\psi}}({^1\!S_0^{[8]}})\rangle,
\langle{\cal O^{J/\psi}}({^3\!S_1^{[8]}})\rangle$
and the data are well described.
Here it is important to note that hadroproduction data alone can not
constrain all three matrix elements even including polarization data.
(ii) The predictions from the NLO NRQCD global analysis
of unpolarized world data do not agree with $J/\psi$ polarization data from CDF I
and the new measurements from CDF II.
With other words it is not possible to describe the unpolarized hadro- and
photoproduction data and the CDF polarization measurements with one set
of CO LDMEs.
Conversely, the new measurements from ALICE agree with NLO NRQCD within errors.
Future precise polarization measurements at the LHC will have the potential
to confirm or dismiss the universality of the LDMEs.

\subsection{Top quark physics}
\label{sec:th_top}

In the past few years, there has been impressive theoretical progress 
in the calculation of top quark pair production beyond NLO.
A highlight has certainly been the completion of the exact 
NNLO QCD corrections to $q\bar{q} \to t \bar{t}+X$ 
which represents a theoretical breakthrough because it is the first 
ever NNLO calculation involving more than two coloured particles
and/or massive fermions \cite{Mitov,Baernreuther:2012ws}. 
As a last missing piece for this calculation,
suitable counter terms needed to regulate
infrared divergences in the interference terms of
tree-level and one-loop amplitudes with three particles
in the final state (of which two are massive) have been 
derived in \cite{Bierenbaum,Bierenbaum:2011gg}.
For $\mu_R = \mu_F =: \mu$ the partonic cross section reads in NNLO ($k\le 2$):
\begin{equation}
\hat{\sigma}_{q\bar{q}}(\beta,m^2,\mu^2) = 
\frac{\alpha_s^2}{m^2} \sum_{k=0}^2 \sum_{l=0}^k \alpha_s^k \sigma_{q\bar{q}}^{(k,l)}(\beta) L^l\, ,
\end{equation}
where $L=\ln(\mu^2/m^2)$. 
The scale-dependent terms $\sigma_{q\bar{q}}^{(k,l)} L^l$ with $l \ge 1$
can be generally computed from the lower order 
functions $\sigma_{q\bar{q}}^{(k'<k,l=0)}$ by renormalization group methods
\cite{Langenfeld:2009wd} so the new information resides in the
function $\sigma_{q\bar{q}}^{(2,0)}(\beta)$. 
Furthermore, it is possible to perform an all order resummation
of universal Sudakov logarithms $\ln \beta$, which become
dominant close to the production threshold ($\beta \to 0$),
at next-to-next-to-leading logarithmic (NNLL) accuracy.
%
The NNLO+NNLL result allows one to predict the $t\bar{t}$ production cross
section at the Tevatron with significantly improved precision 
at the level of 3\% \cite{Baernreuther:2012ws}. 
A C++ program for the calculation of the $t\bar{t}$ total cross section
including the full NNLO corrections is publicly available \cite{Czakon:2011xx}.

In addition to the production threshold ($\beta=\sqrt{1-4 m_t^2/\hat{s}} \to 0$)
all order resummations can also be performed for the 
pair invariant mass (PIM) threshold $z = M_{t\bar{t}}^2/\hat{s} \to 1$ 
treating the $t\bar{t}$ system as a pair
and the one particle inclusive (1PI) threshold $s_4 = \hat{s} +\hat{t}_1 + \hat{u}_1 \to 0$
where one integrates over the phase space of one of the heavy quarks.
The latter two thresholds are relevant for
differential distributions in the pair invariant mass and for the
transverse momentum or rapidity of the observed heavy quark, respectively,
in addition to the total cross section
and different approaches to resum them exist in
the literature \cite{Yang,Kidonakis:2012db,Kidonakis:2011ca}.
The differences between the methods concern formally subleading terms
which, however, can be numerically important.
Alternatively, approximate NNLO theories can be constructed by
combining the information from the exact NLO calculation
with all universal (soft and collinear) logarithmic terms appearing at NNLO, 
see also Sec.\ \ref{sec:th_dis}.

Within the approach based on soft-collinear-effective theory (SCET),
phenomenological predictions for the total cross section \cite{Ahrens:2011px}
and various differential distributions \cite{Ahrens:2010zv,*Ahrens:2011mw}  
at the Tevatron and the LHC 
have been discussed \cite{Yang}.
Furthermore, using standard momentum-space resummation in pQCD,
a large number of numerical results for the differential and total
cross sections for both, 
top pair 
\cite{Kidonakis1} 
and single top production
\cite{Kidonakis2}, 
have been presented by N. Kidonakis \cite{Kidonakis:2012db}.
One observable which has received a lot of attention recently is the 
top quark charge asymmetry. It was shown, that the higher order predictions
are consistent with NLO and hence do not resolve the discrepancy
with the Tevatron data at high invariant mass and rapidity \cite{Ahrens:2011uf}.


More exclusive top observables are important because they
can have sizable production cross sections or constitute 
important backgrounds for Higgs searches.
Here theoretical progress has been achieved by 
merging exact NLO calculations with parton showers. 
At this conference, results have been presented for the production
of top quark pairs with one jet ($t\bar{t}+j$)
\cite{Alioli:2012hj,Alioli:2011as}.
In addition, several processes 
($t\bar{t}+j$ \cite{Kardos:2011qa}, $t\bar{t}+Z$ \cite{Kardos:2011na,*Garzelli:2011is}, 
$t\bar{t}+H/A$ \cite{Garzelli:2011vp} and $W^+W^-b\bar{b}$)
have been implemented into the PowHel framework
as has been discussed in a talk by A.\ Kardos \cite{Kardos}.

\section{Charm and beauty production}
\label{sec:cb}
\subsection{Heavy quarkonium production}
Measurements of the  production cross sections of prompt $J/\psi$
 have been performed by the four LHC collaborations. In addition CMS~\cite{Chatrchyan:2011kc} and LHCb~\cite{Aaij:2012ag} also
presented measurements of $\psi(2S)$ production which is less
influenced by feed-down from the decays of heavier states. These
results are reasonably well described by the NRQCD predictions including
CS+CO contributions at NLO as discussed in the theory
section~\ref{sec:th}. Calculations based on the $k_T$ factorization
approach including only CS diagrams are also able to describe the data.

The situation is less clear when the polarization of the produced state is considered.
The new measurements of the $J/\psi$  and $\Upsilon(nS)$ helicity,
presented respectively by the ALICE~\cite{Abelev:2011md} and
CDF~\cite{CDF:2011ag,Kambeitz} collaborations, agree
marginally with the NRQCD NLO prediction.  The CDF $\Upsilon$ result, obtained
with a integrated luminosity of 5.7~fb$^{-1}$, is in good agreement
with previous CDF data obtained with lower luminosity while the disagreement with the
measurement by the D0 collaboration persists. 

\subsection{Charm production in DIS and $F_2^{\rm c\bar{c}}$}
\begin{figure}[tb]
  \centering
  \begin{minipage}{0.38\textwidth}
  \centering
  \includegraphics[width=0.75\textwidth, clip=true]{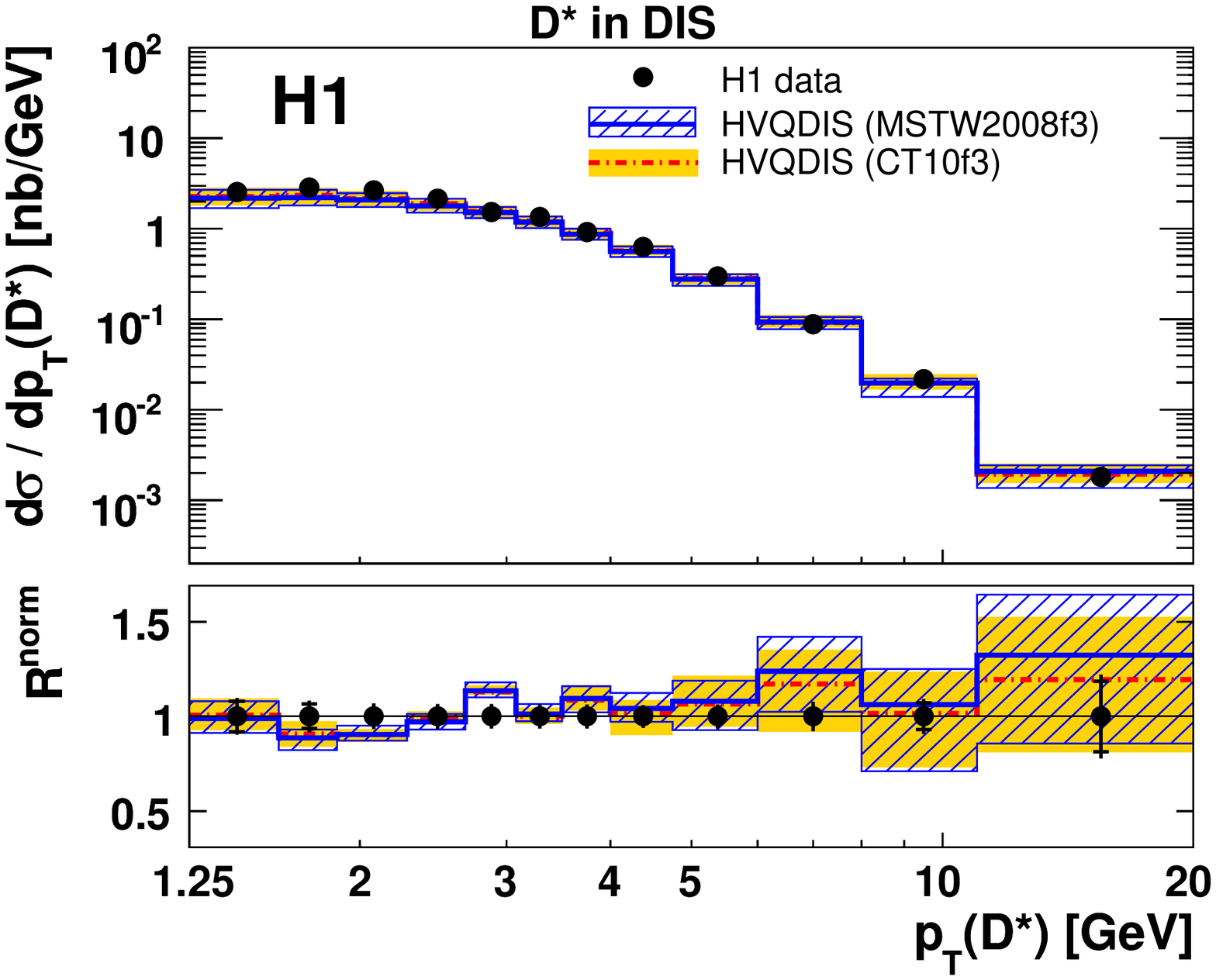}
  \includegraphics[width=0.75\textwidth, clip=true]{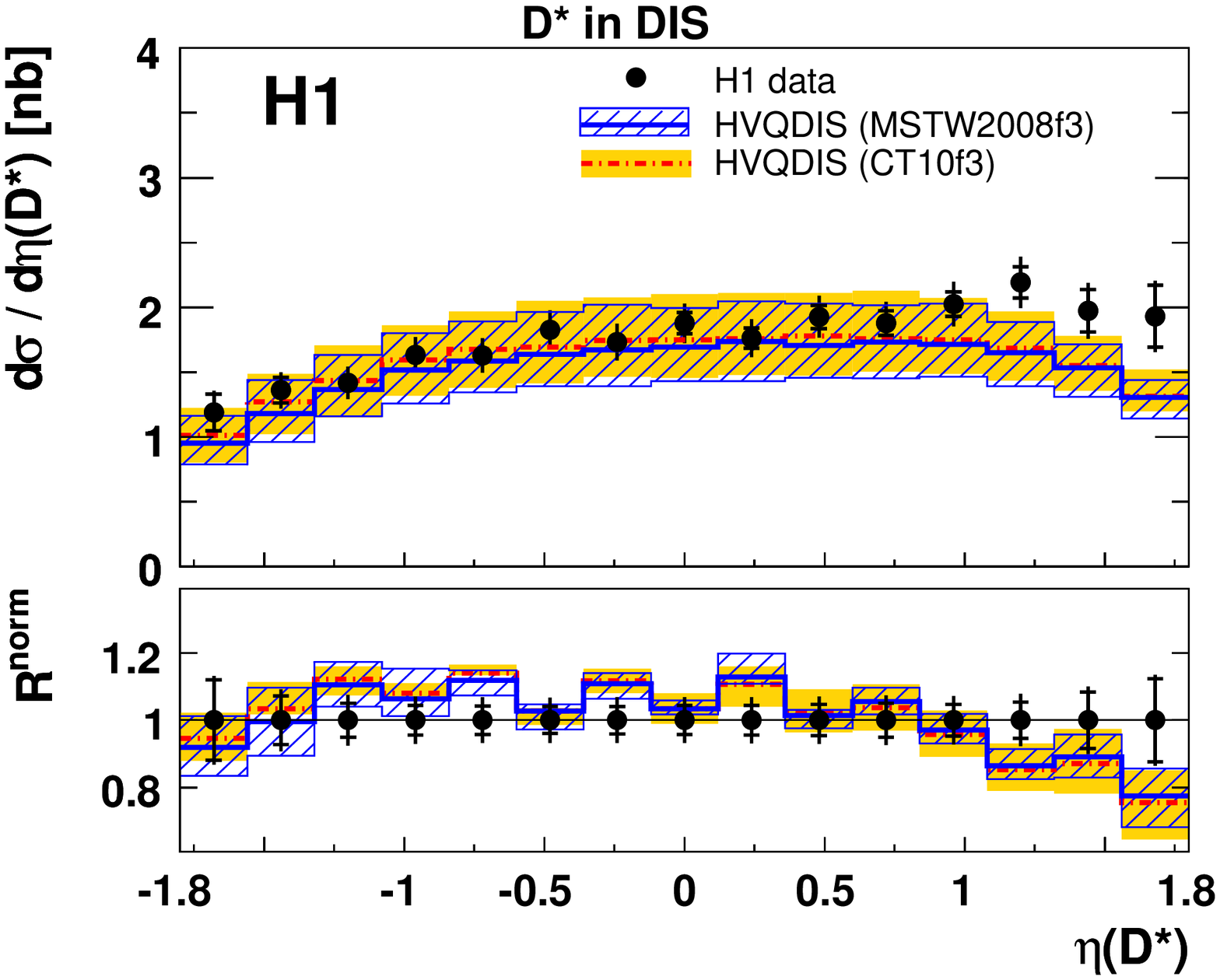}
\end{minipage}\begin{minipage}{0.52\textwidth}
  \centering 
 \includegraphics[width=\textwidth, clip=true]{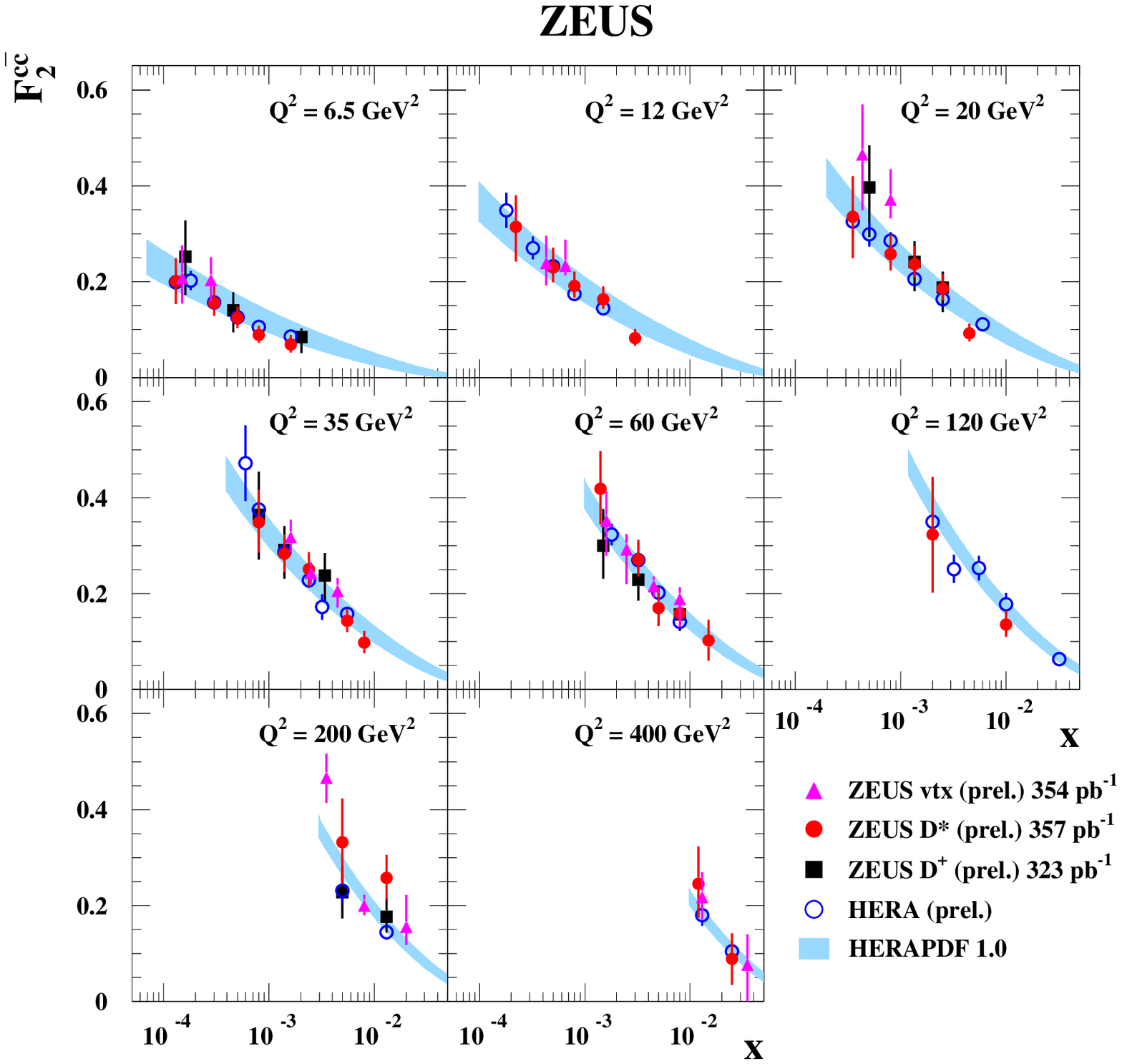}
\end{minipage}
  \caption{Left: H1 measurement of the $D^{*+}$ meson production in DIS 
    for $Q^2>5$~GeV$^2$, $p_T>1.25$~GeV and $|\eta|<1.8$ as a function of
    $p_T$ and $\eta$,  compared to NLO QCD calculations. Right: the charm structure
    function $F_2^{\rm c\bar{c}}$. Recent preliminary ZEUS results
    based on secondary vertex tagging, on $D^{*+}$ and $D^+$ mesons are
    compared to the preliminary combination of H1 and previous ZEUS
    data and to a GM-VFNS prediction based on the HERAPDF1.0 PDFs.}
  \label{f:charmdis}
\end{figure}

The H1 and ZEUS collaborations are finalizing their effort to measure
charm and beauty production in DIS and to provide the best possible
measurement of $F_2^{\rm c\bar{c}}$, the component of the inclusive
structure function $F_2$ 
with charm in the final state.
The H1 collaboration published their final measurement of $D^{*+}$
production in DIS~\cite{Aaron:2011gp}, while ZEUS presented preliminary results on
$D^{*+}$ production and on charmed jets tagged using secondary
vertices~\cite{Gizhko, Libov}. These results are reasonably well described by NLO QCD calculations as
shown  in Fig.~\ref{f:charmdis}~(left) for the $D^{*+}$ case. Charm cross section measurements obtained with
different techniques are used to extract   $F_2^{\rm c\bar{c}}$. Figure
~\ref{f:charmdis}~(right) shows the ZEUS preliminary measurements
compared to a preliminary combination including H1 data and ZEUS
measurements from older data sets.  Once combined, these data will 
improve the knowledge of  $F_2^{\rm c\bar{c}}$, reaching a precision
of about 5\% over a wide range of $x$ and $Q^2$.

\subsection{Charm photo- and hadro-production}
The H1 collaboration presented a new measurement of the $D^{*+}$
photoproduction cross sections~\cite{Aaron:2011mp} that has been compared to various QCD
calculations: fixed-order NLO, the NLO Monte-Carlo matched to 
parton-shower (MC@NLO)~\cite{Frixione:2003ei} and GM-VFNS. They all describe
well the data within their theoretical uncertainties. 
The ZEUS collaboration presented a precise measurement of the charm
fragmentation fractions into different charmed hadrons,
 based on photoproduction data~\cite{Dolinska}. The good agreement with $e^+e^-$
results  supports the universality of charm fragmentation.

\begin{figure}[!h!tbp]
  \centering
  \begin{minipage}{0.49\textwidth}
  \centering
  \includegraphics[width=0.9\textwidth, clip=true]{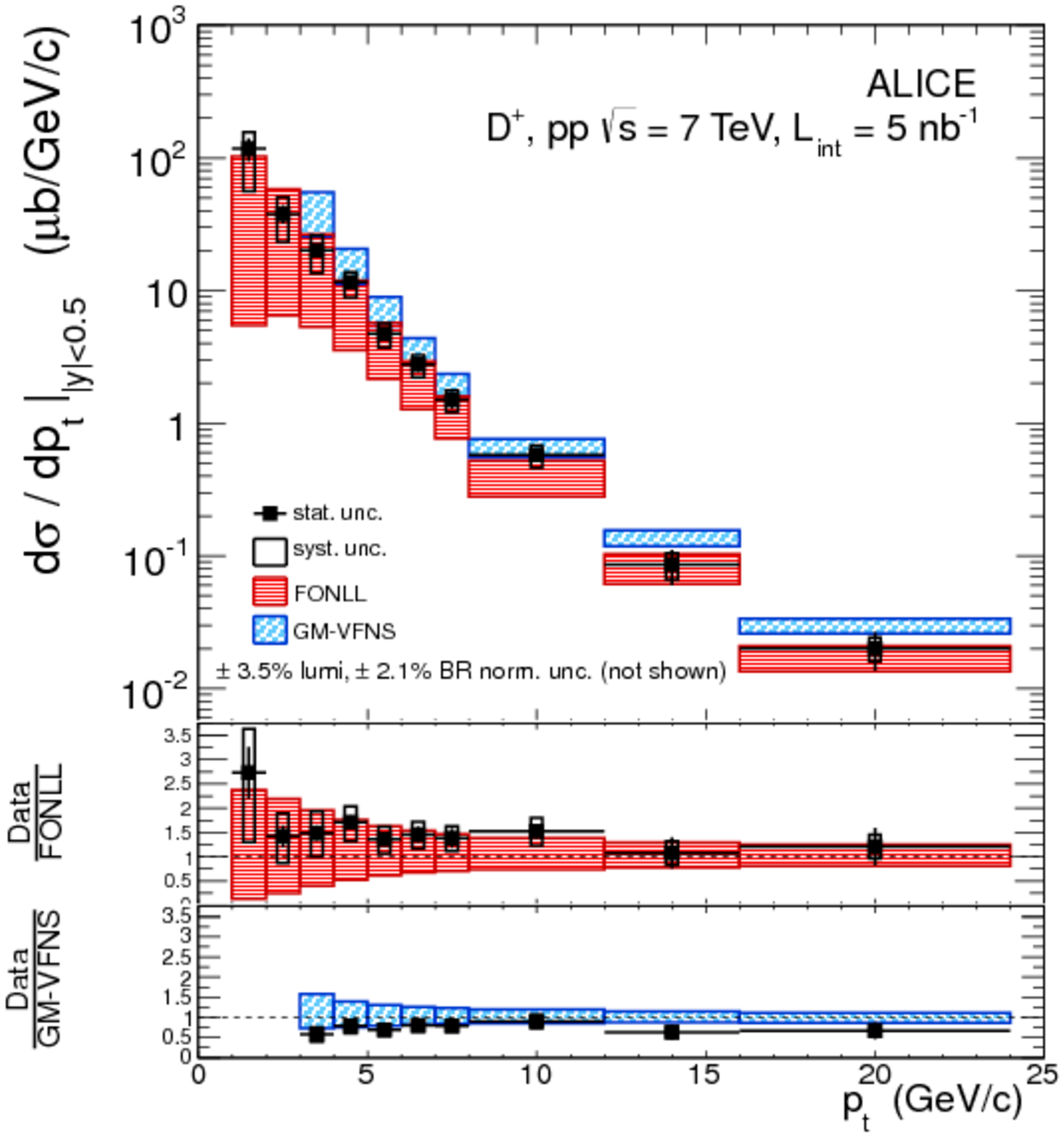}
\end{minipage}\begin{minipage}{0.49\textwidth}
  \centering 
 \includegraphics[width=0.75\textwidth, clip=true]{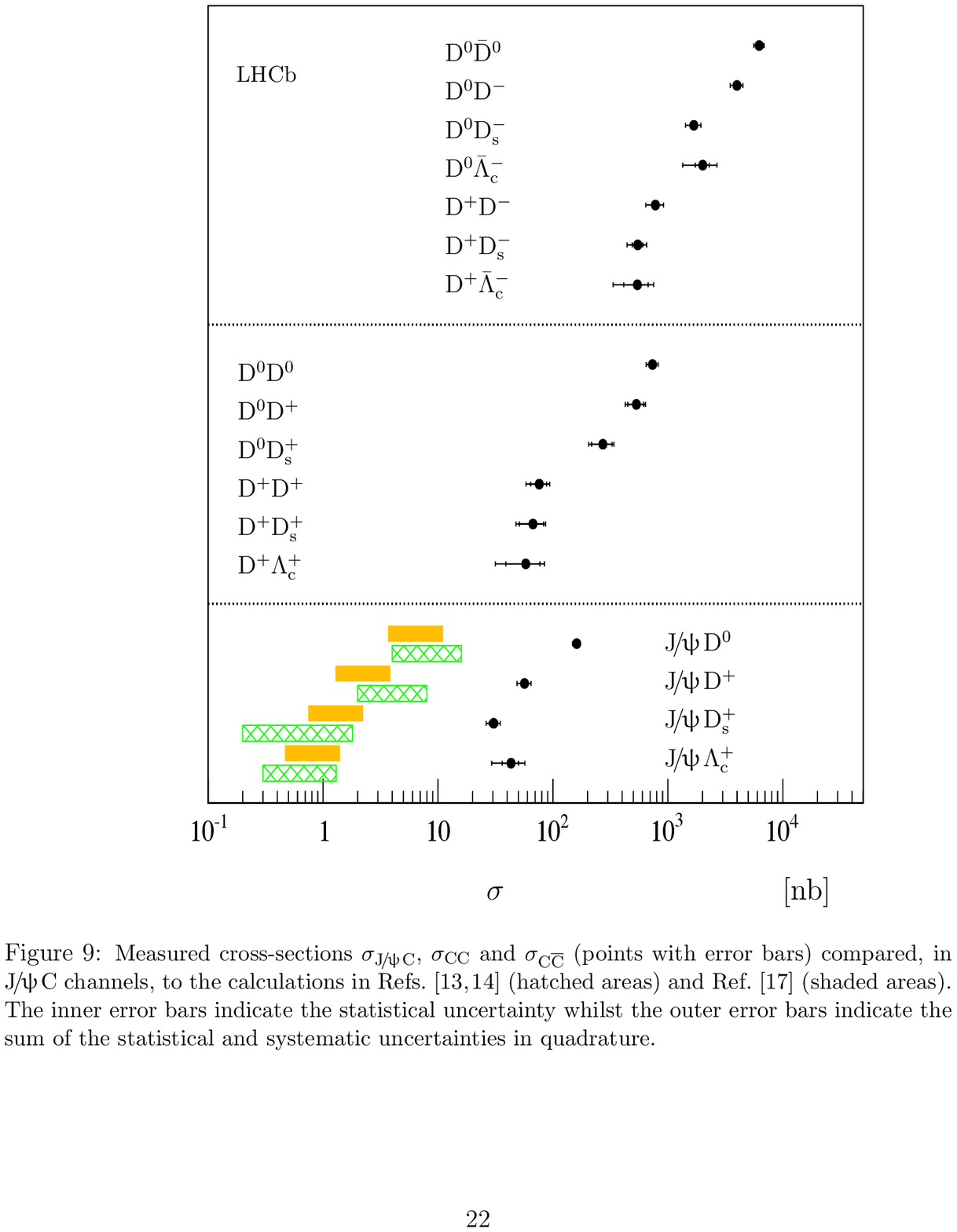}
\end{minipage}
  \caption{Left: $D^+$ meson production cross section
    differential in $p_T$ as measured by ALICE  in pp collisions compared to FONLL and
    GM-VFNS predictions. Right: cross section for the production of two hadrons
    containing charm quarks as measured by LHCb for $2<y<4$ and $3<p_T<12$~GeV: $C \bar{C}$ (up), $CC$ (middle), $J/\psi C$
    (bottom). Filled areas correspond to theoretical calculations for gluon-gluon processes.}
  \label{f:charm}
\end{figure}

The ATLAS and ALICE collaborations presented differential cross sections for the
production of $D$ mesons~\cite{Barton,ALICE:2011aa}. The results are in agreement
with QCD predictions at NLO matched to next-to-leading-log resummation (FONLL)~\cite{Cacciari:1998it}
and, at transverse momenta larger than the charm mass, 
also to GM-VFNS calculations \cite{Spiesberger:2012zn,*gmvfns_lhc}.
Compared to FONLL,
data are in general on the upper edge of the theoretical
uncertainty, which is particularly large at low transverse momenta, as
shown e.g. in Fig.~\ref{f:charm}~(left).
Less inclusive observables do not match completely the standard expectations.
This is the case of the measurement of $D^*$ jets by ATLAS~\cite{Aad:2011td}, in
which the fraction of the jet momentum carried by the associated $D^*$
is on average lower than the predictions based on standard QCD
programs and on the charm fragmentation functions measured in $e^+e^-$
experiments. The measurement of double charm ($CC$) and charmonium-charm ($J/\psi C$) production
made by the LHCb collaboration~\cite{LHCB:2012dz} is also challenging our
understanding of heavy-flavour production. The rate of $CC$ meson
pairs is $\approx 10\%$ of the $C\bar{C}$ rate, as shown in Fig.~\ref{f:charm}~(right). 
Considering that $cc$
production occurs at  $O(\alpha_s^4)$, while $c\bar{c}$ pairs are
produced at $O(\alpha_s^2)$, this result appears unexpected at a first
sight. Anyway a comparison with detailed QCD calculations is needed
to  understand  the origin of this large $CC$ fraction. Proposed explanations are
multiparton interactions or the excitation of intrinsic charm in the proton.

\subsection{Beauty production}
\begin{figure}[!h!tbp]
  \centering
  \begin{minipage}{0.49\textwidth}
  \centering
  \includegraphics[width=0.9\textwidth, angle=90, clip=true]{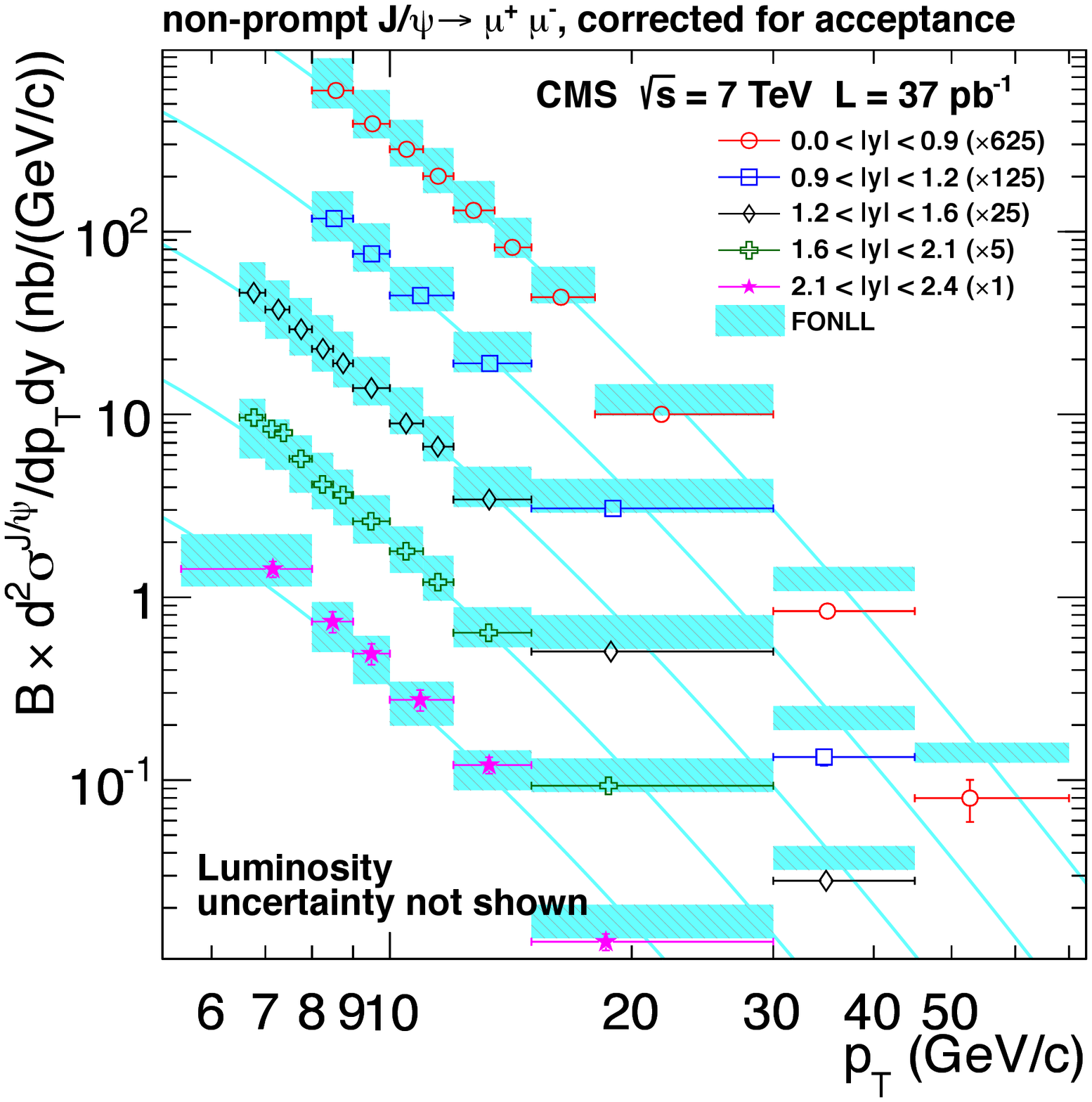}
\end{minipage}\begin{minipage}{0.49\textwidth}
  \centering 
 \includegraphics[width=0.95\textwidth, clip=true]{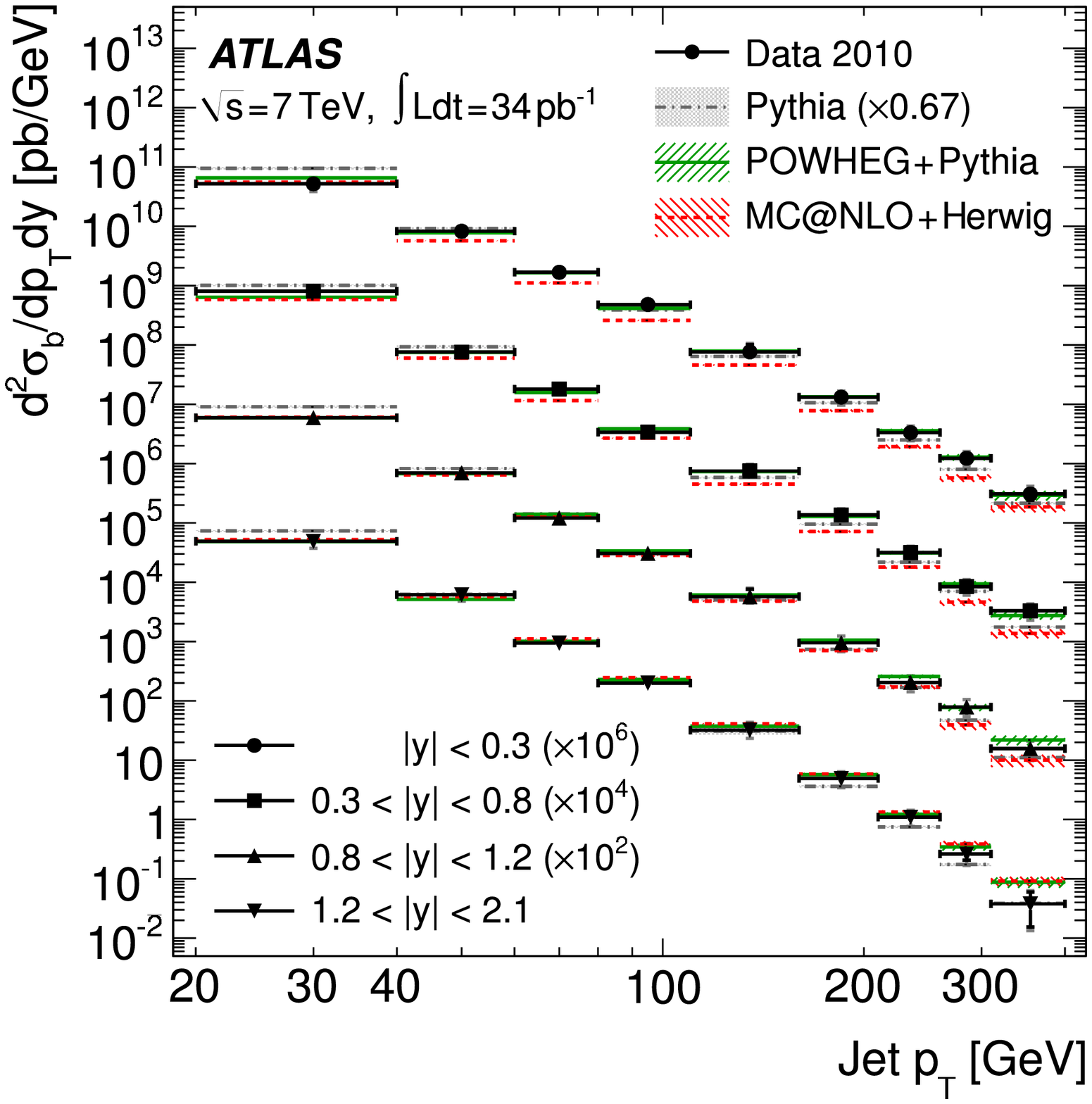}
\end{minipage}
  \caption{Left: non-prompt $J/\psi$ production from b-hadron decays measured by CMS
    compared to FONLL predictions. Right: comparison of the inclusive
    b-jet cross section measured by ATLAS to various QCD predictions.}
  \label{f:b}
\end{figure}

New measurements of open beauty production at the LHC have been presented,
using different techniques in different kinematic ranges.
All the four LHC experiments measured beauty production by
tagging non-prompt $J/\psi$ or $\psi(2S)$ from B-hadron decays, 
covering a large range in $p_T$ (from zero to tens of GeV) and in
rapidity ($0<y <4.5$).  The results are in good agreement with the FONLL theory within
relatively large theoretical uncertainties, as shown in Fig.~\ref{f:b}~(left) for the
CMS case. They CMS collaboration also presented a measurement of the cross section
for di-muons originating from the decays of $b \bar{b}$ pairs~\cite{Chatrchyan:2012hw} which provides precise data
in the low-$p_T$ regime. The measured cross section is in agreement
with the MC@NLO prediction within the $\approx 30\%$ theoretical uncertainty.

At high $p_T$, the measurement of beauty production has been extended
by the ATLAS and CMS collaborations using  b-jet tagging algorithms~\cite{ATLAS:2011ac,Chatrchyan:2012dk}. 
Figure~\ref{f:b}~(right) shows, as an example, the  ATLAS inclusive
b-jet cross section in bins of $p_T$ and rapidity, measured using 2010
data and compared to the predictions from the MC@NLO and the POWEG+Pythia~\cite{Frixione:2007nw} 
NLO Monte Carlos. The agreement is good, especially with POWEG+Pythia
which features a better treatment of beauty fragmentation and decays.
The uncertainty is dominated by the experimental
systematics on b-tagging efficiency  and on b-jet energy scale that
sum up to $\approx 16\%$ at $p_T=100$~GeV and central rapidity.

\section{Top quark physics}
\label{sec:top}
Many top quark measurements have been made at the Tevatron and the LHC,
both in top quark pair and single top production. At the Tevatron, results using
half or all of the final data set have been released, and results from the LHC with 
2011 data collected at 7~TeV are available.
\begin{figure}[!h!tbp]
  \centering
  \includegraphics[width=0.45\textwidth]{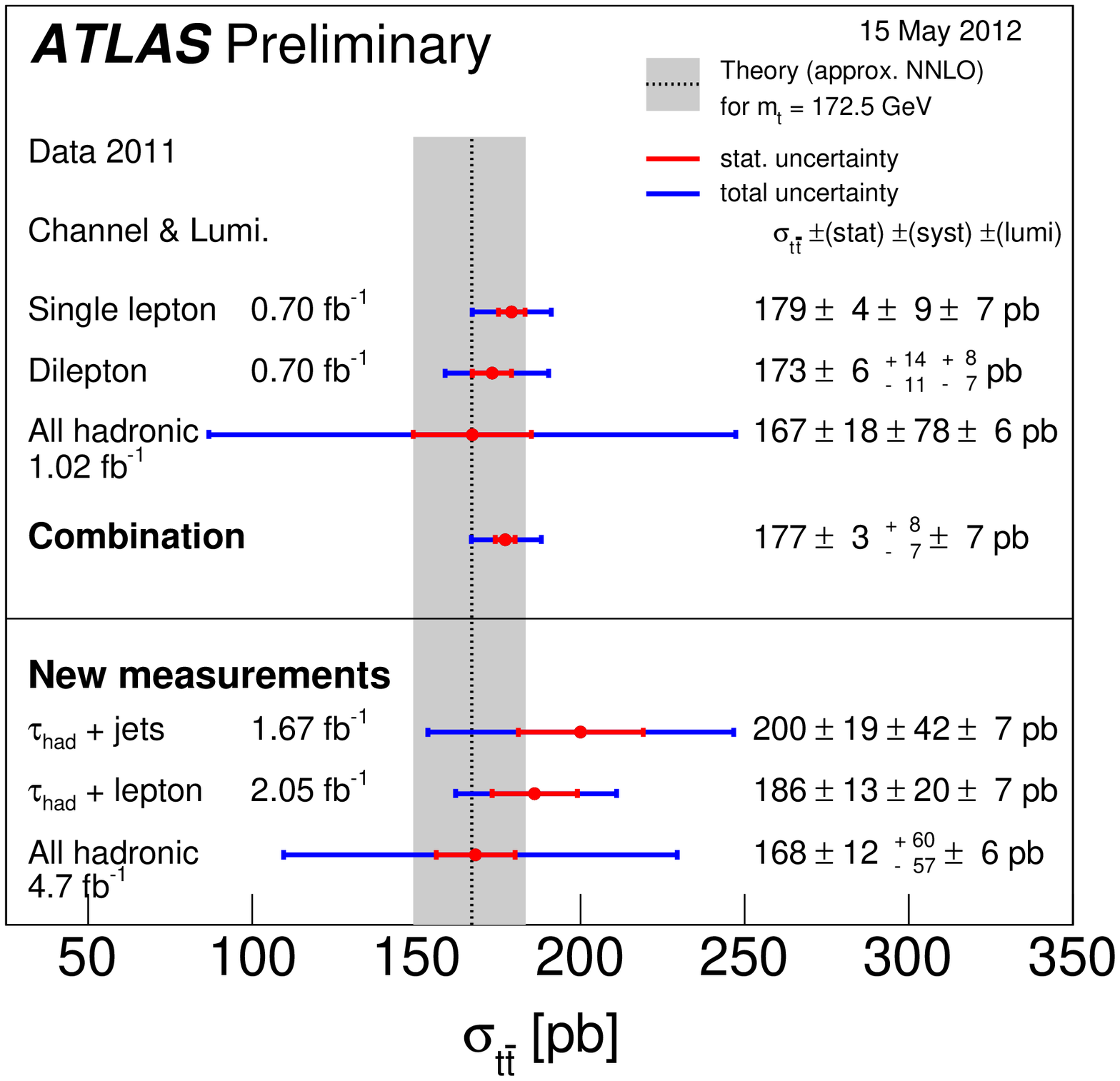}
  \includegraphics[width=0.45\textwidth]{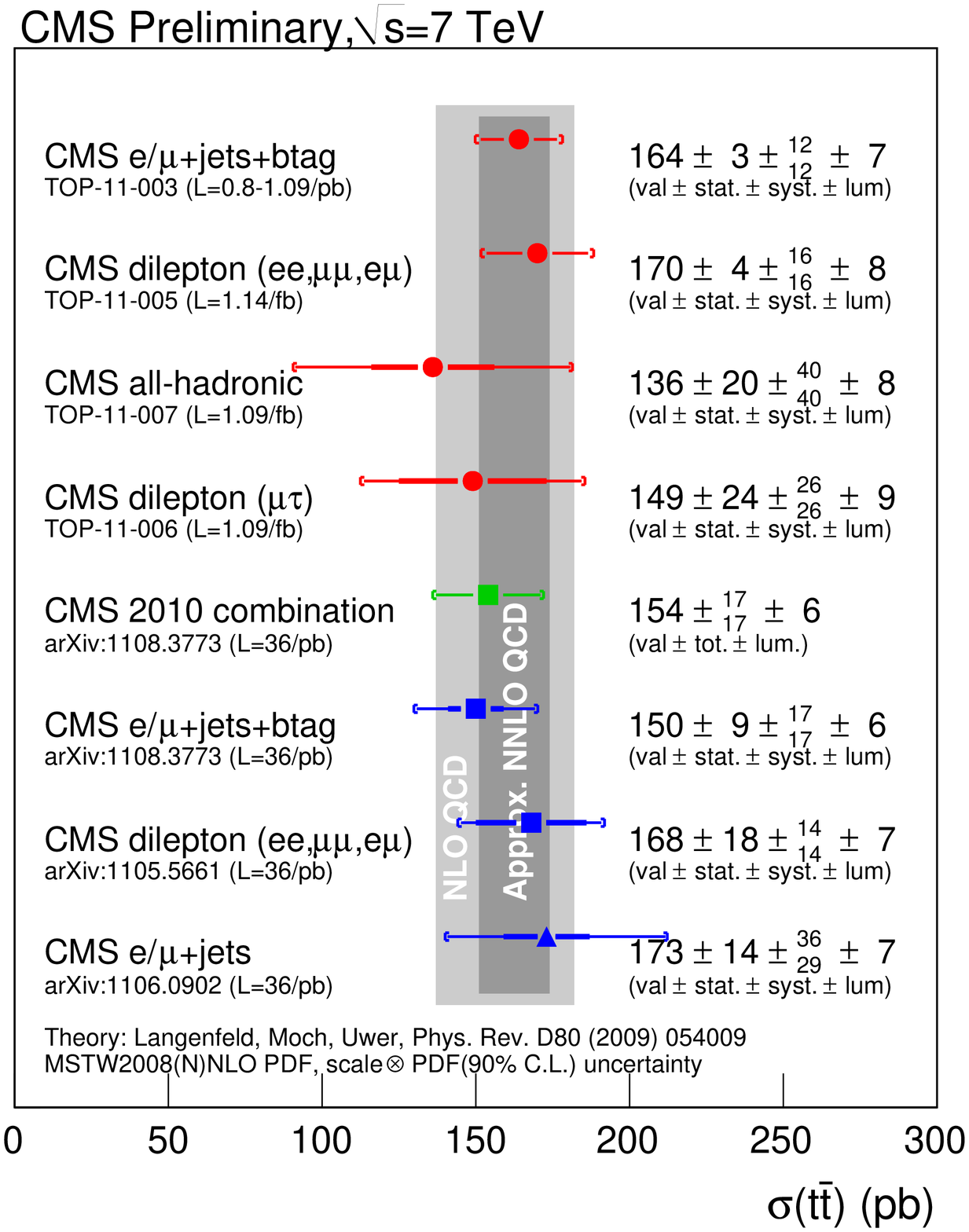}
  \caption{Summary of top quark pair production cross section measurements
  (left) by ATLAS and (right) by CMS.}
  \label{Fig:topxs}
\end{figure}

\subsection{Top quark pair production}
The top quark pair production cross section has been measured in different
final states. Cross section measurements from the CDF and D0 collaborations 
with up to 5~fb$^{-1}$ of 1.96~TeV proton-antiproton data utilize electrons,
muons, taus and all-hadronic final states. Each experiment has
measured the cross section with an uncertainty of better than 
8\%~\cite{Aaltonen:2010ic,Abazov:2011mi}.

The LHC collaborations ATLAS~\cite{Aad:2012xh,ATLAS:2012aa} and 
CMS~\cite{Chatrchyan:2012vs,Chatrchyan:2011yy} have measured the top quark pair 
production cross section in several different final state configurations, 
summarized in Fig.~\ref{Fig:topxs}. This includes tau lepton
final states, and for ATLAS even the tau+jets and all-hadronic final states.
All measurements show good agreement with the SM expectation.

CMS additionally has a preliminary measurement of the differential cross section 
as a function of $p_T$ and rapidity of the 
$t\bar{t}$~system~\cite{CMS-PAS-TOP-11-030}. ATLAS has preliminary measurements of 
the $t\bar{t}$~cross section with a veto on forward jets~\cite{ATLAS:2012al} and 
of the cross section for $t\bar{t}$~production in association with a 
photon~\cite{ATLAS-CONF-2011-153}.

\subsection{Single top quark production}
\begin{figure}[!h!tbp]
  \centering
  \includegraphics[width=0.42\textwidth]{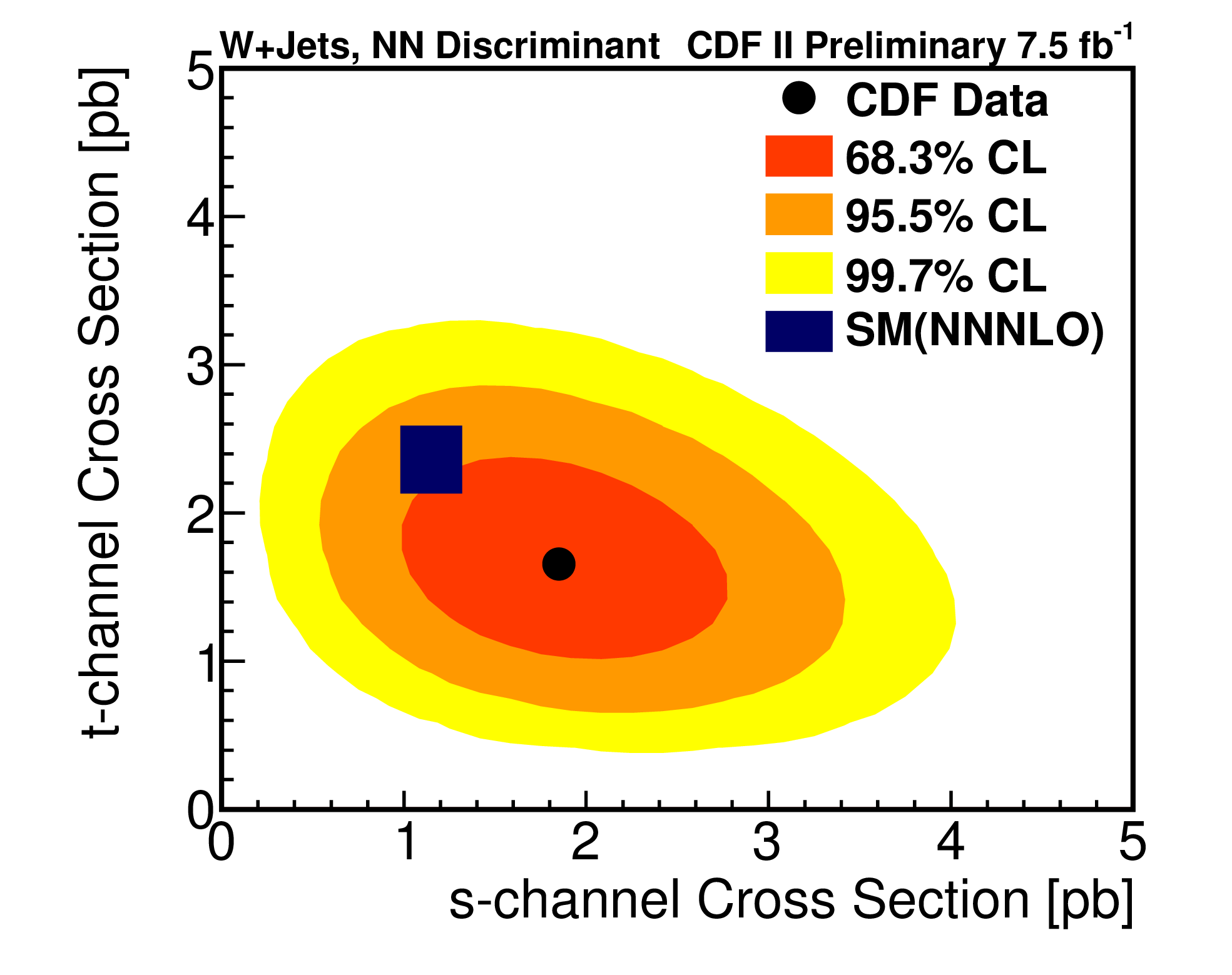}
  \includegraphics[width=0.56\textwidth]{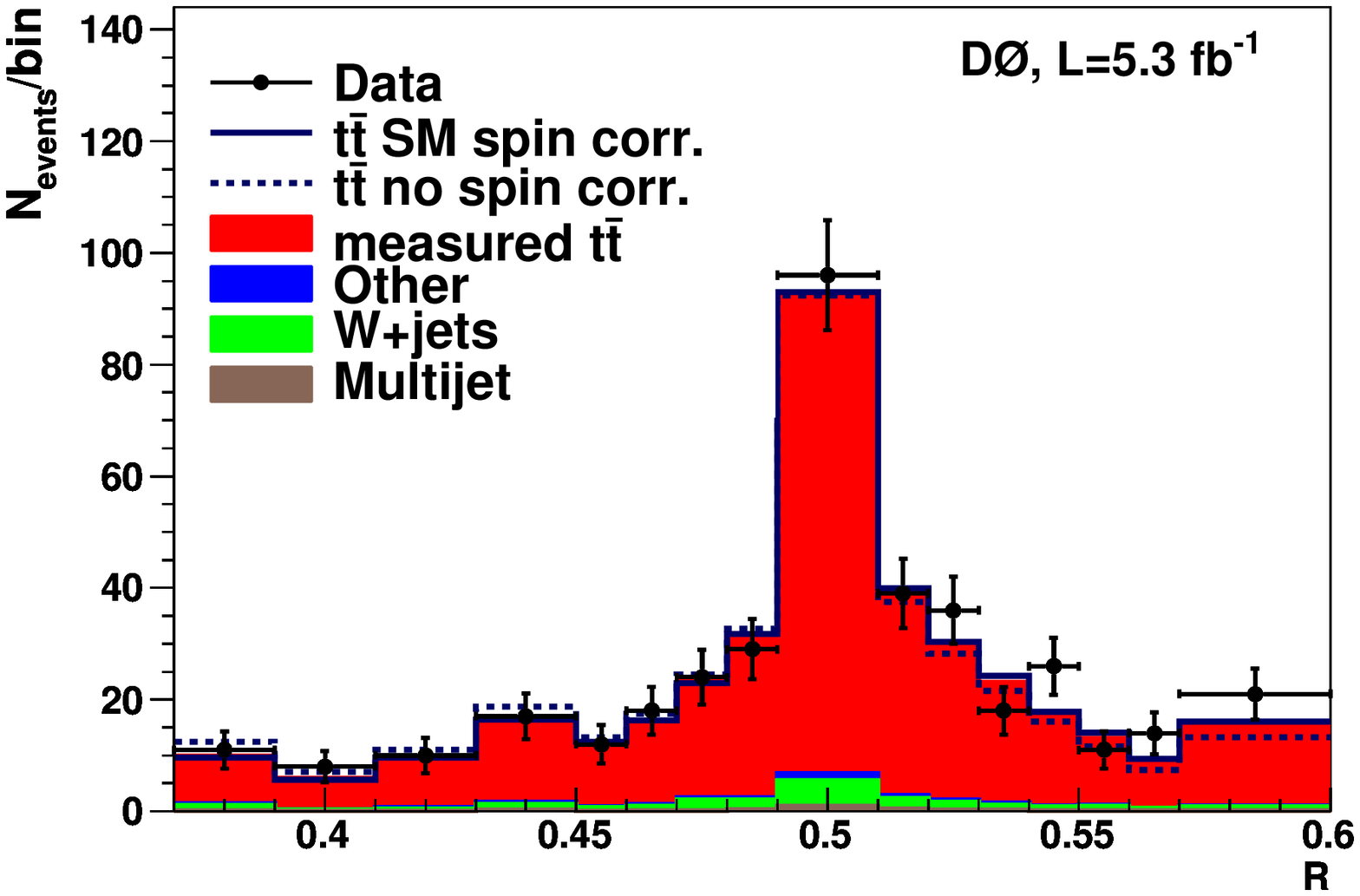}
  \caption{(left) Single top quark production cross section for $s$-channel and 
  $t$-channel from CDF and (right) top quark spin correlation measurement by D0.}
  \label{Fig:singletop}
\end{figure}

Single top quark production has been observed both at the Tevatron and the LHC.
D0 has measured the cross section for the $t$-channel production 
mode~\cite{Abazov:2011rz} as well as the total single top production cross 
section~\cite{Abazov:2011pt}. CDF has a preliminary measurement of single top 
quark production based on 7.5~fb$^{-1}$ of data~\cite{CDF:10793}. 
Figure~\ref{Fig:singletop}~(left) shows the 2d contour of $t$-channel vs
$s$-channel cross sections measured by CDF and compares them to the SM.

ATLAS has measured the $t$-channel cross section to be $83\pm 20$~pb using 
1~fb$^{-1}$ of 7~TeV data using a neural network 
approach~\cite{Collaboration:2012ux}. A cut-based measurement also provides 
separate top and antitop quark $t$-channel cross 
sections~\cite{Collaboration:2012ux}.  CMS has measured a
$t$-channel cross section of $70.3\pm 11.5$~pb~\cite{CMS-PAS-TOP-11-021}. ATLAS and CMS
also have first searches for $Wt$ associated 
production~\cite{ATLAS-CONF-2011-104,CMS-PAS-TOP-11-022}, but have not yet
measured the cross section. ATLAS also has searched for $s$-channel single top
quark production~\cite{ATLAS-CONF-2011-118}.

The single top quark cross section is proportional to the CKM matrix element
$|V_{tb}|^2$, and all collaborations have derived lower limits on $|V_{tb}|$.
D0 has also extracted a measurement of the top quark width from the $t$-channel
cross section through a combination with the flavour composition 
measurement in top quark decays~\cite{Abazov:2012vd}.

The single top quark final state is sensitive to many models of new physics. 
Recent searches for flavor-changing neutral currents~\cite{Aad:2012gd}
and new heavy bosons by ATLAS~\cite{Aad:2012ej}, and for anomalous couplings by 
D0~\cite{Abazov:2011pm,Collaboration:2012iw} have not found any evidence for
new physics and set stringent limits.

\subsection{Top quark mass}
The mass of the top quark has been measured with high precision at both the
Tevatron and the LHC. D0 has a new top quark mass measurement in the dilepton
channel~\cite{Abazov:2012rp} and CDF has a new measurement in the lepton+jets
channel~\cite{CDF:10761} with an uncertainty of 1.3~GeV. The Tevatron average
of 173.2~GeV has an uncertainty of only 1.0~GeV~\cite{Lancaster:2011wr}.

ATLAS and CMS have not yet reached that level of precision but have also produced
first results. The ATLAS top quark mass measurement has an uncertainty of 2.4~GeV
and the CMS measurement has an uncertainty of 1.3~GeV.
CMS also measures the mass
difference between top and antitop quarks~\cite{Chatrchyan:2012ub}.

\subsection{Top quark properties}
Abundant top quark samples are now available making measurements such as spin
correlation in top quark pair production possible. D0 has reported evidence
for spin correlation~\cite{Abazov:2011gi}, shown in 
Fig.~\ref{Fig:singletop}~(right).  A preliminary CDF study finds a value 
consistent with no spin correlation in a similar size data set~\cite{CDF:10719}.
ATLAS also observes spin correlation~\cite{ATLAS:2012ao}.

Since the Tevatron is a proton-antiproton collider, a forward/backward asymmetry
in top quark pair production can be measured in the rapidity difference between
the top and antitop quark. CDF has a preliminary result for
the full Tevatron data set~\cite{CDF:10807}, finding a significant deviation from
the SM. D0 also sees a deviation from the SM expectation~\cite{Abazov:2011rq}.

At the LHC proton-proton collider, a possible asymmetry is reflected as a charge
asymmetry which is more difficult to measure. The ATLAS measurement of
the charge asymmetry is consistent with the SM 
expectation~\cite{ATLAS:2012an}. The CMS measurement is also consistent with
the SM~\cite{Chatrchyan:2011hk}.

\section{Spectroscopy and rare decays}
\label{sec:other}


\subsection{Spectroscopy of heavy quarkonia and b-hadrons}

Comprehensive study of heavy quarkonia is an important test of the QCD. 
The goal is to complete the predicted $c \bar{c}$ and $b\bar{b}$ spectra, 
measure their properties and transitions. Such an experimental information 
allows validating the QCD predictions obtained either in the theoretical 
approach, where the $Q\bar{Q}$ multiplets are calculated on the lattice, 
or in the phenomenological one i.e.~from the potential models which attempt 
to model QCD features by describing the interquark potential. 

In the charmonium landscape, all the states lying below the threshold for 
decays to open charm were discovered. Agreement with predictions of the potential 
models is quite good, whereas the lattice usually underestimates some 
of the splittings, for instance the $J/\psi-\eta_c(1S)$ hyperfine one~\cite{qwg-report}. 
The era of precise lattice QCD calculations for charmonia above 
the $D\bar{D}$ threshold has only begun. There are only a few such charmonia 
observed, whereas most of the missing states (except for ground $D$-wave states) 
are expected to decay dominantly into $D^{(*)}\bar{D}^{(*)}$-like final states 
and to have quite large widths. 
Over the last decade many $c\bar{c}$-like states, called the $X,\ Y,\ Z$, 
were observed by experiments at $e^+e^-$ as well as hadron colliders~\cite{qwg-report}. 
Their properties are either unusual for conventional charmonia (large widths for 
hadronic transitions, non-zero electric charge) or simply don't match any empty 
slots in the $c\bar{c}$ spectrum. Therefore they are considered as candidates for 
exotic hadrons, like molecules, tetraquarks or hybrids, which are also predicted 
within the QCD framework. The experiments continue their efforts to confirm and$/$or 
further study properties of these resonances. Belle have investigated 
the properties of the most famous $c\bar{c}$-like state, the $X(3872)$ in its 
discovery decay mode $J/\psi \pi^+ \pi^-$. The world best upper limit on its width was 
set at $1.2~\textrm{MeV}/\textrm{c}^2$~\cite{x3872_belle}. Neither the charged partner 
$X^{\pm} \to J/\psi \pi^{\pm} \pi^0$ nor the $C$-odd partner searched for in 
$J/\psi \eta$ and $\chi_{c1} \gamma$ final states have been observed. 
Instead, Belle found the first evidence of the narrow $\psi_2(1D)$ charmonium 
decaying to the $\chi_{c1} \gamma$ at the mass of 
$3823 \pm 3~\textrm{MeV}/\textrm{c}^2$~\cite{other_belle}. Babar have reported 
study of the $\eta_c \pi^+ \pi^-$ produced in two-photon annihilation. 
Such a study is important to test an interpretation of the $X(3872)$ as 
the $\eta_{c2}(1D)$ or to search for the $\eta_{c2}(1D)$ itself, but no significant 
signal for neither of them has been found. 
Charged $c\bar{c}$-like states play a special role, as they must consist of 
at least four quarks. Belle observed three such resonances,  
the $Z(4430)^{\pm} \to \psi' \pi^{\pm}$ and $Z(4050)^{\pm}, Z(4250)^{\pm} 
\to \chi_{c1}\pi^{\pm}$ produced in $B \to Z^{\pm} K$ decays~\cite{z_belle}, 
however Babar have not confirmed any of them~\cite{z_babar}. Since the upper 
limits set by Babar on the product branching fractions do not contradict 
the Belle measurements, the conclusive results are expected to come from 
the LHC experiments. Both LHCb and CMS have demonstrated that they will 
play an important role in studies of $c\bar{c}$ spectroscopy. 
Using only a small fraction of their data, they measured precisely mass 
and $pp$ production of the $X(3872)$~\cite{x3872_lhcb,x3872_cms}.
The $X(3872)\to J/\psi \pi^+ \pi^-$ yield expected with the 2011 Run data
will hopefully allow them to discriminate between the two possible spin-parities 
of the $X(3872)$, $1^{++}$ and $2^{-+}$.

Experimental data on the bottomonium spectrum remain even more incomplete. The field 
has become lively once $B$-factories took data at $\Upsilon(\textrm{n} S)$'s   
other than $\Upsilon(4S)$, which allowed Babar to discover the ground bottomonium 
$\eta_b(1S)$ in the $\Upsilon(2,3S) \to \eta_b(1S) \gamma$ transitions~\cite{etab_babar}. 
Belle studies of the data collected at the $\Upsilon(5S)$ have revealed that 
its properties differ from other $\Upsilon(\textrm{n} S)$'s. A reason for abnormally 
large transitions $\Upsilon(5S) \to \Upsilon (1, 2, 3 S) \pi^+ \pi^-$~\cite{yb_belle} 
are two charged $b\bar{b}$-like states 
$Z_b(10610)^{\pm}, Z_b(10650)^{\pm} \to \Upsilon (1,2,3S) \pi^{\pm}$ that mediate these 
transitions~\cite{zb_belle}. The $\Upsilon(5S) \to h_b(1,2P) \pi^+ \pi^-$ transitions 
have been found to be as large and also saturated with 
the $Z_b(10610)^{\pm}, Z_b(10650)^{\pm} \to h_b(1,2P) \pi^{\pm}$ amplitudes. 
Interpretation of the $Z_b^{\pm}$ states as $B^{(*)}\bar{B}^{*}$ molecules 
seems to be supported by their masses and decay amplitude pattern~\cite{zb}. 
The $\Upsilon(5S) \to h_b(1,2P) \pi^+ \pi^-$ transitions allowed the first 
observation of the spin-singlet $h_b(1P)$ and $h_b(2P)$ bottomonia~\cite{belle_hb}. 
Belle have also reported a measurement of the mass and the first measurement of the width of 
the $\eta_b(1S)$ produced in $h_b(1P) \to \eta_b(1S) \gamma$. Measured 
$\eta_b(1S)-\Upsilon(1S)$ hyperfine splitting improved agreement with the 
theoretical predictions~\cite{qwg-report}. The $b\bar{b}$ spectroscopy is also 
studied at the hadron colliders. The first particle observed at the LHC has 
been a candidate for $\chi_b(3P)$ found at $10530 \pm 51~\textrm{MeV}/\textrm{c}^2$ 
by Atlas in $\chi_b(3P) \to \Upsilon(1,2S) \gamma$ transitions~\cite{chib_atlas}; 
it has been soon confirmed by D\O~\cite{chib_d0}. 
 
LHCb has taken over a leading role in studies of $b$-hadrons being out of reach of 
the $B$-Factories. The presented results on $B_c$ mesons, such as a mass measurement using 
the $B_c^+ \to J/\psi \pi^+$ decays, their production relative to topologically similar 
$B^+ \to J/\psi K^+$ reference decays, as well as the first observation of the 
$B_c^+ \to J/\psi \pi^+ \pi^- \pi^+$ decays, comprise an important 
experimental input on this heaviest $b$-meson observed so far~\cite{bhadrons_lhcb}. 
The world best mass measurements of $b$-baryons: $\Lambda_b^0$, $\Xi_b^-$ and 
$\Omega_b^-$ and study of the $\Lambda_b$ production allowing extraction of 
the $\frac{f_{\Lambda_b}}{f_d+f_u}$ production fraction, nicely improve previous 
Tevatron measurements. 

\subsection{Rare decays of heavy flavours}

Search for physics beyond the Standard Model (SM) is nowadays one of the main 
goals of particle physics. Precision studies of the $B_{(s)}$, $D_{(s)}$ and $\tau$ 
decays which in the SM are either suppressed or forbidden, allow one to search 
for New Physics (NP) effects. Virtual contributions of NP particles can enter 
the diagrams underlying the studied processes and modify the SM predictions for various 
observables, like decay rate, CP violation or polarization. The deviation of 
a given observable from the SM will be a sign of NP, whereas the study of correlations 
between observables will allow one to identify the nature of NP. Such indirect 
searches are complementary to the direct ones performed at the LHC. 
All the measurements performed so far to overconstrain the unitarity triangle 
(UT) describing $B$ meson system seem to be consistent with each other within 
their uncertainties~\cite{ckmfitter} and thus constrain NP corrections to be 
at most at the $10 \%$ level. Numerous tensions between the measurements and the 
SM predictions reported over the last few years encourage to investigate them 
further with increased precision. 

The asymmetry of same-sign muon pairs from semileptonic $B_{(s)}$ decays was found 
by D$\O$ to be larger than the SM prediction by $3.9 \sigma$~\cite{dimuons_d0}.  
This asymmetry is related to $CP$ violation in $B^0-\bar{B}^0$ and 
$B^0_{s}-\bar{B}^0_{s}$ mixing. A phase entering the latter, the $\phi_s$, 
can be independently measured via a time dependent analysis of $B^0_s \to J/\psi \phi$, 
where $CP$ violation occurs via interference between the decay proceeding with 
and without $B_s$ mixing. Updated measurements of the $\phi_s$ and $\Delta \Gamma_s$ 
by CDF~\cite{phis_cdf} and LHCb~\cite{phis_lhcb} show that the overall agreement with 
the SM is good. This might suggest the NP contribution in the $B^0-\bar{B}^0$ mixing, 
and such a picture seems to be supported by a $2.5 \sigma$ discrepancy between 
the $\sin {2 \beta}$ UT parameter measured directly from the 
$B^0 \to c \bar{c} K^0$ decays and indirectly determined from the global fit 
to all the remaining UT measurements~\cite{ckmfitter}. 

The golden mode for NP searches, the $B^0_s \to \mu^+ \mu^-$, is very clean from 
the theoretical as well as the experimental point of view. The SM prediction for its
branching ratio is only $(3.2 \pm 0.2) \times 10^{-9}$, while a broad class 
of NP models can enhance it up to $10^{-7}$. 
A global effort has been made to find the $B^0_s \to \mu^+ \mu^-$ signal 
and the strongest constraint 
of ${\cal BR}(B^0_s \to \mu^+ \mu^-)<3.8 \times 10^{-9}$ comes from 
LHCb~\cite{bs2mumu_lhcb}. CDF measurement updated with their full data 
sample~\cite{bs2mumu_cdf} has not reinforced previously found signal increase.  

Contrary to the $D^0-\bar{D}^0$ mixing, where any NP 
effects would be obscured by long distance contributions, $CP$ 
violation in $D$ meson decays has been over last the years suggested to 
be a good place for NP searches. In the SM it is expected to be very small, 
at the level of $10^{-3}$ at most, and experimental measurements have reached 
that precision only recently. First evidence of CP violation in charm decays 
has been found by LHCb in the measurement of 
$\Delta A_{CP}=A_{CP}(D0\to K^+K^-)-A_{CP}(D0\to \pi^+\pi^-)=(-0.82 \pm 0.21 \pm 0.11)\%$
~\cite{cpv_d_lhcb}, which in a first approximation corresponds to the difference 
of direct CP violation. Together with the result reported by CDF~\cite{cpv_d_cdf}, 
no CP violation scenario is excluded at the $4 \sigma$ level. However, according 
to the revised theoretical calculations such an increased CP violation can be still 
accommodated within the SM~\cite{cpv_d_theory}.  

Indirect searches for NP in the heavy flavour sector have been very lively. 
Many rare processes only come within reach thanks to increasing 
sensitivity obtained with data from LHC and full data samples from $B$-Factories 
and Tevatron experiments being analyzed. Some of the important studies, like 
precision ${\cal BR}(B \to \tau \nu)$,  polarization measurements 
in $B \to \bar{D}^{(*)} \tau \nu$, lepton flavour violation in $\tau$ decays will 
be feasible only with Super$B$-Factories being under construction. 

\section{Conclusions}


There has been theoretical progress concerning approximate NNLO calculations
for heavy quark structure functions in deep inelastic scattering,
the extension of the ACOT heavy flavour scheme to jet production,
and advances in top physics (soft gluon resummations, merging with parton showers) 
where the highlight is clearly the first complete NNLO QCD prediction for top pair 
production in the $q \bar{q}$ annihilation channel.
Furthermore, state of the art phenomenological predictions for 
open charm and bottom, charmonium, and single top
and top pair production have been discussed in addition to other topics
such as the effect of double parton scattering on heavy quark production.

A huge amount of measurements of heavy-quarkonium and of open charm and beauty production
has been produced from HERA, LHC, Tevatron, and RHIC.  These measurements challenge the
QCD predictions that are typically less precise than
experimental data. The agreement between data and theory is in general good. 
Some measurements, for which  the agreement is
marginal,  deserve further studies:
the polarization of heavy quarkonium, the measurement of jets
associated  to a $D^{*+}$ presented by ATLAS, and the measurement of double charm
production performed by LHCb.

The top quark is being scrutinized with unparalleled precision at
both the Tevatron and the LHC. The D0 and CDF experiments are measuring the
strong and electroweak production cross section as well as the top quark mass
very precisely. 

New $c\bar{c}$ and $b\bar{b}$ states have been observed and exotic quarkonia
extensively studied by $B$-Factories as well as LHC and Tevatron experiments.
LHCb have performed the world best measurements of properties of $B_c$ and
$b$-baryons.

Rare decays of $D_{(s)}$, $B_{(s)}$ and $\tau$ have been measured
with increased sensitivity. Some of them show tensions with Standard
Model predictions and thus give hints of New Physics.


{\raggedright
\begin{footnotesize}




\end{footnotesize}
}


\end{document}